\documentclass[12pt,reqno]{article}

\usepackage{epsfig}
\usepackage{amstex}

\begin{document}

\begin{titlepage}
\hfill TUW--96--08 \\
\hfill gr-qc/9605044 \\
\begin{center}

{\LARGE Hawking radiation and masses in generalized dilaton theories}\\ 
\vfill
\renewcommand{\baselinestretch}{1}

{\large
{H. Liebl \footnote{liebl@@tph.tuwien.ac.at}}}\\
Institut f\"ur Theoretische Physik \\
Technische Universit\"at Wien\\
Wiedner Hauptstr. 8-10, A-1040 Wien\\
Austria \\[4ex]

{\large
{D.V. Vassilevich \footnote{vasilevich@@phim.niif.spb.su} and S. Alexandrov}}\\
Department of Theoretical Physics\\
St. Petersburg University\\
198904 St. Petersburg \\
 Russia
\end{center}
\vfill

\begin{abstract}
A generalized dilaton action is considered of which the standard dilaton
black hole and spherically 
reduced gravity are particular cases.
The Arnowitt-Deser-Misner (ADM) and the Bondi-Sachs (BS) mass are calculated.
Special attention is paid to both the asymptotic conditions for the metric
as well as for the reference space-time. For the latter one we suggest a modified
expression thereby obtaining a new definition of energy. Depending on the 
parameters of the model the Hawking radiation behaves like a positive or negative
power of the mass.

\end{abstract}
\end{titlepage}
\vfill

\section{Introduction}

Over the last few years 1+1 dimensional dilaton theories have been studied 
extensively in their string inspired (CGHS) version \cite{cghs} as well as in more
general forms \cite{dilat}. One of the main motivations to study such models 
arises from the hope that within the simplified setting of 2D models one can
gain insight into physical properties of 4D gravity.
They actually allow for example black hole solutions and Hawking radiation 
and are more amenable to quantum treatments than their 4D counterparts.
The two most frequently considered theories, the string inspired CGHS 
as well as spherically
reduced gravity (SRG) differ drastically in some of their physical properties.
For example differences were observed with respect to 
the completeness of null geodesics for 
those 
two models \cite{liebl}. These differences directly lead one to investigate
physical properties of a generalized 
model of which the two prominent examples are simply particular cases.

Important classical quantities are the energy at spatial and null infinity,
the so called Arnowitt-Deser-Misner (ADM) and the Bondi-Sachs (BS) mass, 
respectively
(provided the global structure of the space-time is the same as for SRG).
For general models, however, there exist in general no flat space time solutions
\cite{liebl} and therefore a proper reference space-time has to be chosen.
A second important point is to realize that these ``masses'' depend
on the asymptotic behavior of the metric and are therefore only defined
with respect to a particular observer.
Our approach is based on the  second order formalism. Significant insight 
can also be gained from a first order formalism \cite{strobl},\cite{kummer}.

An important feature in semiclassical considerations
is the behavior of Hawking radiation. In the CGHS model
it is just proportional to the cosmological constant whereas the dependence 
in SRG is inverse to its mass, which implies an accelerated evaporation
towards the end of its lifetime.
As we will show a generalized theory will
exhibit Hawking radiation which is proportional to the black hole mass in terms of
positive or negative powers of the black hole mass, depending on the parameters 
of the model.

In section 2 we repeat some relevant results of \cite{liebl}
for the metric and the global
structure of the classical solution of a generalized dilaton Lagrangian which,
for a certain range of parameters posesses a singularity structure coinciding
with the one of the Schwarzschild black hole. 
The ADM and BS mass are calculated in section 3.
Furthermore we will present a definition for energy, differing from the one 
proposed by Hawking \cite{hawking}, by taking a modified reference space-time.
The path integral
measure and the problem of interpretation of various energy definitions will
be discussed in section 4 before we finally demonstrate the dynamical formation
of black holes in the frame work of conformal gauge, with special emphasis
on the necessary boundary conditions.

\section{Classical Solution}

Among the numerous different generalizations of the 
CGHS model \cite{cghs}
we consider the action
\begin{equation}
\label{ldil}
L=\int d^2x \sqrt{-g}e^{-2\phi}(R+4a(\nabla\phi)^2 +Be^{2(1-a-b)\phi}).
\end{equation}
This form of the Lagrangian covers e.g. the CGHS model \cite{cghs}
for $a=1$, $b=0$, spherically 
reduced gravity \cite{lau} $a=\frac{1}{2}$, $b=-\frac{1}{2}$, 
the Jackiw-Teitelboim model \cite{jackiw} $a=0$, $b=1$.
Lemos and Sa \cite{lemos} give the global 
solutions for $b=1-a$ and all values of $a$,  
Mignemi \cite{mignemi} considers
$a=1$ and all values of $b$. The models of \cite{fabbri} correspond to $b = 
0, a \leq 1$.

The classical solutions of (\ref{ldil}) were already found in \cite{liebl} but
since the solutions for the metric will be used extensively in the following 
sections they shall be repeated here.
Letting $\phi$ represent one of the coordinates the line element reads

\begin{equation}
\label{metric}
(ds)^2= g(\phi) \left( 2dvd\phi+ l(\phi) dv^2 \right),
\end{equation}
with
\begin{eqnarray}
\label{oldg}
{} & g(\phi)= e^{-2(1-a)\phi} \\
\label{Gl:34}
b \neq -1: & l(\phi)=\frac{e^{2\phi}}{8} \left(C- \frac{2B}{b+1} e^{-2(b+1)\phi} \right), \\
\label{Gl:34b}
b =-1: & l(\phi)=\frac{e^{2\phi}}{8} \left( \tilde{C}+4B\phi \right),~~~\tilde{C}=C-2B \ln 2 .
\end{eqnarray}
The transformation 
\begin{eqnarray}
\label{Gl:34c}
a\neq1: &u=\frac{e^{-2(1-a)\phi}}{2(a-1)}, \\ a=1: &u=\phi,
\end{eqnarray}
brings the metric into Eddington-Finkelstein form (EF) \cite{ef} 
\begin{equation}
\label{ds2}
(ds)^2= 2dvdu+l(u)dv^2
\end{equation}
with
\begin{eqnarray}
\label{newl}
b\neq -1: & a\neq 1: & l(u)=B_1|u|^{\frac{a}{a-1}} -B_2|u|^{1-\frac{b}{a-1}}, \\
 & a=1: & l(u)=\frac18 e^{2u}\left(C-\frac{2B}{b+1}e^{-2(b+1)u} \right), \\
b=-1: & a \neq 1: & l(u)=\frac{1}{8}|2(a-1)u|^{\frac {a}{a-1}}
 \left(C+\frac{2B}{a-1}\ln|2(a-1)u|\right), \\
 & a=1: & l(u)=\frac{e^{2u}}{8} \left( \tilde{C} +4Bu \right),
\end{eqnarray}
where the constants are given by
\begin{equation}
\label{a1}
B_1=\frac{C}{8} (2|a-1|)^{\frac{a}{a-1}} \qquad B_2 = \frac{B}{4(b+1)}(2|a-1|)^
{1-\frac{b}{a-1}}.
\end{equation}
The scalar curvature of the metric (\ref{metric}) has the form
\begin{eqnarray}
b\ne -1 & \quad & R=\frac 12 e^{2(2-a)\phi} \left ( aC
-\frac {2Bb}{b+1} (b+1-a) e^{-2(1+b)\phi} \right ) \nonumber \\
b=-1  & \quad & R=\frac 12 e^{2(2-a)\phi} \left ( aC
+2(a+1)B +4aB\phi \right ) \label{curva}
\end{eqnarray}
or equivalently for (\ref{newl})
\begin{eqnarray}                                        \label{curv}
b\ne-1, & a\ne1: & R=B'_1u^{\frac{2-a}{a-1}}+B'_2u^{\frac{a+b-1}{1-a}}
\\                                                      \label{ecnboa}
b\ne-1, & a=1: & R=\frac C2e^{2u}-\frac{Bb^2}{b+1}e^{-2bu}
\\                                                      \label{ecobna}
b=-1, & a\ne1: & R=|2(a-1)u|^{\frac{2-a}{a-1}}\left(\frac{aC}{2}+
B\frac{a^2-1+a\ln|2(a-1)u|}{a-1} \right)
\\                                                      \label{ecoboa}
b=-1, & a=1: & R=\frac12e^{2u}(\tilde{C}+4B+4Bu).
\end{eqnarray}
Finally, a third form of the metric, the generalized Schwarzschild metric,
\begin{equation}
\label{genschw}
(ds)^2=l(u)dt^2-\frac{1}{l(u)}du^2,
\end{equation}
is obtained by means of the transformation
\begin{equation}
\label{ef2gs}
dv=dt-\frac{du}{l(u)}.
\end{equation}
Since below we will repeatedly make use of the special cases of asymptotically 
Minkowski, Rindler and de Sitter space-times we list them 
for completeness. After a suitable rescaling (see eq.(\ref{K})below) they are
\begin{align} 
\raggedleft
\label{dss}
\nonumber {}&{\mbox{\bf Asymptotically Minkowski space:} \quad (b=a-1)}  \\
{}& \qquad (ds)^2=(MU^{\frac{a}{a-1}}-1)dt^2
-(MU^{\frac{a}{a-1}}-1)^{-1}dU^2   \\
\nonumber {}&{\mbox{\bf Asymptotically Rindler space:} \quad (b=0)}     \\
{}& \qquad (ds)^2=(MU^{\frac{a}{a-1}}-B_2^{\frac 12}U)dt^2
-(MU^{\frac{a}{a-1}}-B_2^{\frac 12}U)^{-1}dU^2   \\
\nonumber {}&{\mbox{\bf Asymptotically de Sitter space:} \quad (b=1-a)}   \\
{}& \qquad (ds)^2=(MU^{\frac{a}{a-1}}-B_2U^2)
dt^2-(MU^{\frac{a}{a-1}}-B_2U^2)^{-1}dU^2 \label{desit}
\end{align}
with $M=B_1B_2^{\frac{2-a}{2(a-1)}}$.
As can be seen from (\ref{curv}) and (\ref{dss}) for 
$0<a<2$ these metrics go to the unit metric or to de
Sitter metric at the asymptotic flat
or constant curvature regions. For $a\not\in (0,2)$
we cannot construct static metrics with proper asymptotic
behavior. Throughout this paper asymptotically Minkowski,
Rindler or de Sitter solution will imply $0<a<2$.

The global structure is completely determined by $l(u)$ or equivalently by 
$l(\phi)$, which is nothing else but the norm of the Killing vector whose
zeros correspond to the horizons.
The corresponding 
Penrose diagrams for all possible values of the parameters of
the action (\ref{ldil}) were presented in \cite{liebl}, where it was shown that
genuine Schwarzschild like behavior is restricted to the region $b \leq 0$
and $a < 1.$

\section{ADM and BS Mass}

\subsection{ADM Mass}
\label{admsubsection}

The aim of this section is to calculate the ADM mass for our model.
More precisely, we will obtain the value of a generalized definition
for the energy at spatial infinity, for space-times whose lapse does not
approach $1$ asymptotically.
The standard approach to obtain this quantity is to split off the dynamical
part $h_{\mu \nu}$ and $\rho$ from the metric 
$g_{\mu \nu}=h_{\mu \nu}+\eta_{\mu \nu}$ and the dilaton field
$\phi=\phi_{vac}+\psi$, respectively
and then calculate the energy defined as the value of the standard Hamiltonian
taken on the shell of zero constraints \cite{iofa} \cite{kim}.
Instead of going through these lengthy arguments we shall first review
some recent very elegant methods that arise from the canonical formalism of
gravity \cite{lau} \cite{brown} and will later on apply them to solutions of
our model.
We review what kind of slicing we have to work with and
will then outline the main steps on how to obtain the expression for the 
total energy. 

\begin{figure}[h]
\centering\epsfig{file=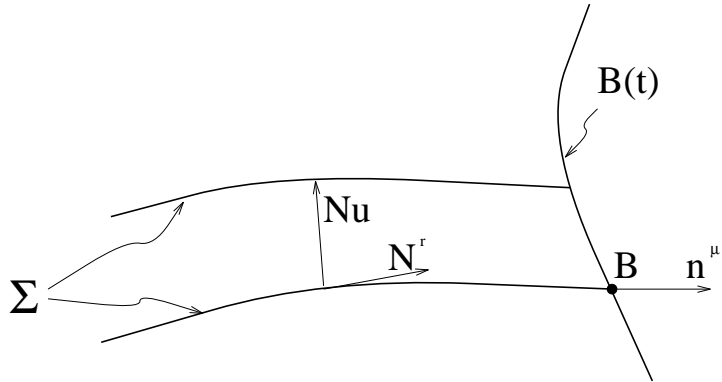,height=5cm}
\caption{Spacetime foliation: $\Lambda^t$ denoting the radial shift and 
$\Lambda$ the radial lapse} \label{adm1}
\end{figure}

Consider a one-dimensional spacelike slice
$\Sigma$ drawn in our spacetime (see Fig.\ref{adm1}). Assume that
$\Sigma$ has a boundary point $B = 
\partial\Sigma$. Let the unit, outward-pointing,
spacelike normal of the point $B$ as embedded 
in $\Sigma$ be $n^{\mu}$. 
If $\Sigma$ is a surface of constant $t$ with metric $\Lambda^2dr^2$
then the space-time metric near $\Sigma$ can be written in ADM form 
\cite{arnowitt}
\begin{equation}
\label{admform}
ds^2=\Lambda^2 dr^2 -N^2(dt+\Lambda^{t}dr)^2.
\end{equation}

To obtain the expression for the total energy one has to cast the action into
hamiltonian form. Boundary terms have to be added to the action to ensure that
its associated variational principle fixes the induced metric and the dilaton
on the boundary. 
As shown in
\cite{lau}, the form of a suitable Hamiltonian
with boundary terms at $B$ is the following:
\begin{equation}
\label{hamil}
H = \int_{\Sigma} {\rm d} r (N {\cal H} +
N^{r} {\cal H}_{r}) + N(E_{ql} + N^{r} 
\Lambda P_{\Lambda})|_{B} ,
\end{equation}
where $P_{\Lambda}$ is the ADM momentum conjugate to $\Lambda$
and ${\cal H}$ and ${\cal H}_{r}$ are the Hamiltonian and the momentum constraint 
respectively. 
Since the expression for the Hamiltonian (\ref{hamil}) diverges in general, a 
reference Hamiltonian $H_0$ has to be subtracted to obtain the physical 
Hamiltonian.  
$E$, which defines the quasilocal energy is given by
\begin{equation}
\label{eql}
E_{ql} = e^{-2\phi}(n[\phi] - n[\phi]^0) 
\end{equation}
where the second term indicates that
the value is referenced to a background.
We shall only consider the case when $N^{r} = 0$ 
at $B$, i.e the on-shell value of Hamiltonian
will be associated only with time translations
($H$ does not generate displacements normal to the
boundary). 
Defining the total energy as the value of the physical hamiltonian on 
the boundary we finally get
\begin{equation}
\label{energy}
E \equiv H|_{B}=NE_{ql}=4Ne^{-2\phi}(n[\phi] - n[\phi]^0)|_{B}
\end{equation}
Only for $N=1$ this coincides with the ADM mass, whereas the additional factor 
$N$ in (\ref{energy}) gives the proper definition for the total energy for 
space-times whose lapse doesn't go to one asymptotically \cite{abbott,hawking}. 
However, as we shall see at the end of this section, the form of (\ref{energy})
is not unique.

To apply this result to our action
(\ref{ldil}) we take the line element
\begin{equation}
\label{pap63}
(ds)^2=g(\phi)l(\phi)dt^2-\frac{g(\phi)}{l(\phi)}d\phi^2
\end{equation}
with the same $g(\phi)$ and $l(\phi)$ as in (\ref{oldg}) and (\ref{Gl:34})
, respectively.
This form of the metric is obtained from (\ref{metric}) by a transformation
analogous to (\ref{ef2gs}).
For $g=1$ ($a=1$) this is just the metric in generalized Schwarzschild coordinates.
We see that $\phi$ now corresponds to the radial coordinate of the ADM-metric 
(\ref{admform}), $\phi \equiv r$.

Since the radial shift  is zero the unit normal $n$ to the boundary $B$
is therefore simply
\begin{equation}
\label{pap63a}
n = \frac{1}{\Lambda}
\left( \frac{\partial}{\partial \phi} \right).
\end{equation}
Using 
\begin{equation}
\label{shift}
\Lambda^2 =-\frac{g(\phi)}{l(\phi)} \quad N^2=-g(\phi)l(\phi)
\end{equation}
we obtain
\begin{equation}
\label{fenergy}
E_{ADM}=4e^{-2\phi} \left.\left(l(\phi)-\sqrt{l(\phi)l^0(\phi)} \right)\right|_{\phi \to \infty}
\end{equation}
where $l^0$ denotes the function in the metric (\ref{metric}) of a
reference space-time. Where possible we will use flat space-time
as our reference. It should be noted that this expression is independent of
the function $g(\phi)$ and therefore we have no $a$-dependence on $E$
for the action (\ref{ldil}).
Calculating this quantity at spacelike infinity corresponds to the ADM mass.
Inserting (\ref{Gl:34}) into (\ref{fenergy}) with $C=0$ for $l^0$ the diverging
terms cancel each other whereas the next order terms give a finite contribution.
We obtain
\begin{eqnarray}
                                                                \label{e1}
E=\frac{C}{4}+O(e^{2(b+1)\phi}) &\qquad &\mbox{for} \qquad(b+1)\phi \to -\infty \\
                                                                \label{e2}
E=\frac{C}{2}+O(e^{-(b+1)\phi}) &\qquad &\mbox{for} \qquad(b+1)\phi \to +\infty.
\end{eqnarray}
This coincides with the expectation that the parameter $C$ is proportional
to the mass.
Notice that for these expressions the reference space-time $(C=0)$ itself
contains already a curvature singularity \cite{liebl} except for the
special cases considered below. For spherically reduced gravity $(a=-b=\frac{1}{2}$
the flat space-time region is located at $\phi \to -\infty$. Therefore, by using
(\ref{Gl:34c}) it is straightforwardly verified that the next order term of
(\ref{e1}) becomes $O(u^{-1})$ which coincides with the usual ADM expression
for the Schwarzschild black hole. 

However, the results (\ref{e1}) and (\ref{e2}) still retain some arbitrariness.
To show this consider the transformation of (\ref{ds2})
\begin{equation}
\label{K}
dV =  K^{\frac{1}{2}}dv \qquad dU = K^{-\frac{1}{2}}du.
\end{equation}
This preserves the form of the metric 
\begin{equation}
\label{pres}
ds^2=2dUdV+L(U(u))dV^2 \qquad L(U(u))=\frac{l(u)}{K}
\end{equation}
and at the same time we are able to obtain dimensions of length for $U$ and $V$.
Simultaneously $E_{ADM}$ is changed. We see that an unambiguous definition
of $E_{ADM}$ includes two important ingredients. The first one is the
reference point for energy which is specified by the choice of ground state
solution. We define the ground state configurations by $C=0$. For the important
particular cases of asymptotically Minkowski , Rindler 
and de Sitter models this choice gives zero or constant
curvature ground state solutions. The second ingredient is the asymptotic
condition for the metric. It corresponds to the choice of an observer
who measures energy and Hawking radiation at the asymptotic region.
This ambiguity in the choice of time for the asymptotic observer is expressed
in (\ref{pres}). In the spirit of these remarks, the value (\ref{e1})
and (\ref{e2}) 
of the ADM mass should be specified as "the ADM mass with respect to $C=0$
solution measured by an asymptotic observer with time and length
scales defined by metric (\ref{metric})". This value is unambiguously
defined for all values of $a$ and $b$. For the important particular cases 
mentioned above other definitions are more relevant: \newline
\labelitemi $\quad$ {\bf Asymptotically Minkowski and Rindler spaces:} \newline
As mentioned above, eq. (\ref{K}) can be used to obtain dimensions of length
for $U$ and $V$. This happens to be the case if $K$ has dimensions 
of energy squared.
Therefore a combination of $B_1$ and $B_2$, e.g. $B_1^mB_2^{1-m}$,
or equivalently of $C$ and $B$, is a natural choice 
since all of these quantities have dimension of energy squared. 
However, because our vacuum is defined as $C=0$, i.e. $B_1=0$, only
$B_2$ should contribute to $K$. If we require in addition that 
for the minkowskian case our metric
approaches asymptotically the unit one $\eta_{\mu\nu}$=diag(-1,1), we have
to use 
\begin{equation}
\label{b2transf}
K=B_{2}
\end{equation}
which in turn gives
\begin{equation}
\label{L}
L(U)=B_{1}B_2^{\frac{2-a}{2(a-1)}}U^{\frac{a}{a-1}}-
B_2^{\frac{a-1-b}{2(a-1)}}U^{1-\frac{b}{a-1}}.
\end{equation}
With
\begin{eqnarray}
n=\frac{1}{\Lambda} \frac{\partial}{\partial U}=
\frac{\sqrt{K}}{\Lambda g(\phi)}\frac{\partial}{\partial \phi} \\
N^2=\Lambda^{-2}=-L(U)
\end{eqnarray}
we obtain from (\ref{energy}) 
\begin{equation}
\label{finaladm}
E_{ADM}=\frac{C}{2} \left( \frac{b+1}{B} \right)^{\frac{1}{2}}
(2(a-1))^{\frac{b-a+1}{2(a-1)}}.
\end{equation}
For the special case of asymptotically Minkowski and Rindler space-times this gives
\begin{equation}
\label{admmink}
E_{ADM}=\frac{C}{2} \left( \frac{a}{B} \right)^{\frac{1}{2}} 
\quad \mbox{and} \quad
E_{ADM}=\frac{C}{2} \left(2B(a-1)\right)^{-\frac{1}{2}},
\end{equation}
respectively.
\newline
\labelitemi $\quad$ {\bf Asymptotically de Sitter spaces:} \newline
A rescaling of (\ref{ds2}) as in (\ref{K}) does not change the asymptotic behavior
of (\ref{genschw}), which is given by 
\begin{equation}
(ds)^2=-B_2u^2dt^2+(B_2u^2)^{-1}du^2,
\end{equation}
but the coefficient of the asymptotically vanishing term is changed
as can be seen from (\ref{L}). 
Therefore we get
\begin{equation}
E_{ADM}=\frac{C}{4(a-1)}\left(\frac{2-a}{B}\right)^{\frac 12}.
\end{equation}
\labelitemi $\quad$ {\bf Modified Definition of Mass:} \newline
Finally we want to present a modification to 
Hawkings definitions for energy.
Eq.(\ref{energy}) was obtained by subtracting the value of the reference space-time
$\left(NE_{ql}\right)^0|_B$ from the Hamiltonian (\ref{hamil}).
Following the line of argument in ref.\cite{hawking} $N$ was pulled
out of the subtraction term by setting $N|_B=N^0|_B$.
However, this condition is not easily implemented because $N$ diverges
in general at the boundary. 
Here we suggest the more natural choice of leaving the lapse in the reference
term which then still fulfills the foregoing condition.
As a consequence this modifies
(\ref{energy}) and (\ref{fenergy}) to
\begin{equation}
\label {newE}
E=4e^{-2\phi}\left.\left(Nn[\phi] - N^0n[\phi]^0\right)\right|_{B}=
4e^{-2\phi}\left.\left(l(\phi)-l^0(\phi)\right)\right|_B
\end{equation}
and by inserting $l(\phi)$ we obtain
\begin{equation}
\label{Eours}
E=\frac{C}{2},
\end{equation}
which now does not only hold for $|\phi| \to \infty$ but on the whole space-time. This 
value depends again in the same manner as above on the asymptotic condition
for the metric.
Note that for asymptotically Minkowski, Rindler or de Sitter solutions
the two expressions (\ref{e2}) and (\ref{Eours}) for $E$ agree. 
On the other hand, for arbitrary asymptotic
behavior the two values of $E$ correspond to two different definitions
of the asymptotic observer in reference space-time. 
In a recent paper \cite{Mann} Mann also considered a modified reference space-time
for the energy to obtain the so called quasilocal mass, which also
diverges from Hawking's proposal. 
Note that the same rescaling as in (\ref{K}) could certainly be applied to 
(\ref{newE}) thereby giving additional coefficients of $B$ and the parameters.
A similar result was also obtained in \cite{kumwid} by applying a Regge-Teitelboim
argument.

\subsection{Bondi-Sachs Mass} 
\begin{figure}[h]
\centering\epsfig{file=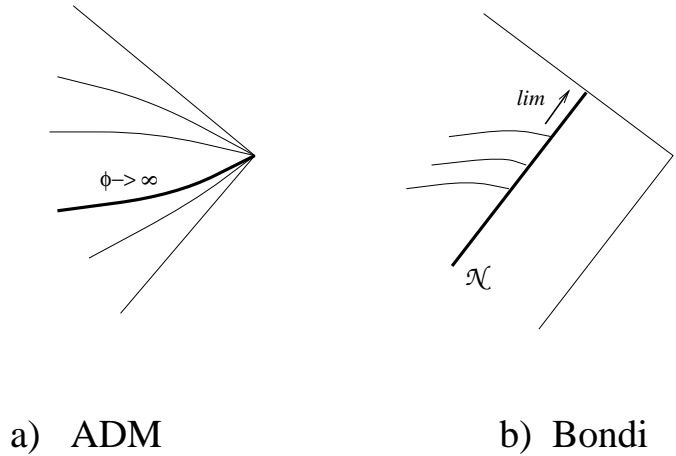,height=6cm}
\caption{Conceptual differences in the construction of the ADM and the Bondi mass}
\label{bondi1}
\end{figure}

In the first part of this chapter we obtained the value of the total
energy which corresponds to the ADM mass at spatial infinity. Here we will
show that $E$ on ${\cal {I}^+}$, the so called BS mass $E_{BS}$,
equals
$E_{ADM}$ and does not depend on the value of the 
retarded time coordinate $v$.
However, the procedure to obtain this quantity is logically different. The main 
difference with respect
to the calculation of the ADM mass above is that we do {\it not}
take the limit $\phi \to \infty$ along a {\it single} slice
but instead we will consider the limit of expression (\ref{energy})
along a particular null hypersurface {\cal N} associated with {\it different} 
slices along a line of constant retarded time \cite{lau2} (see Fig.\ref{bondi1}).
For convenience we will work again with the EF-metric (\ref{ds2}). 
First we define a future directed lightlike vector $k^{\mu}$ and pick another 
lightlike vector $h^{\mu}$ such that $h_{\mu}k^{\mu}=1$. They are
\begin{eqnarray}
k^{\mu}\partial / \partial x^{\mu}&=&\sqrt{\frac{l}{2}}\partial / \partial u   \\
h^{\mu}\partial / \partial x^{\mu}&=&-\sqrt{\frac{l}{2}}\partial / \partial u + 
\sqrt{\frac{2}{l}}\partial / \partial v.
\end{eqnarray} 
Next, along the null hypersurface {\cal N} we let
\begin{eqnarray}
\tilde{u}^{\mu}\partial / \partial x^{\mu} &:=& \frac{1}{\sqrt{2}}
\left( k^{\mu}+h^{\mu} \right) \\
 &=& \frac {1}{\sqrt{l}} \partial /\partial v
\end{eqnarray}
define the timelike normal to the $\Sigma$ slices spanning the points of {\cal N}.
Similarly we define its spacelike normal
\begin{eqnarray}
\tilde{n}^{\mu}\partial / \partial x^{\mu}&:=&\frac{1}{\sqrt{2}}
\left(k^{\mu}-h^{\mu}\right) \\
&=& \sqrt{l} \partial /\partial u - \frac{1}{\sqrt{l}} \partial /\partial v.
\end{eqnarray}

\begin{figure}[h]
\centering\epsfig{file=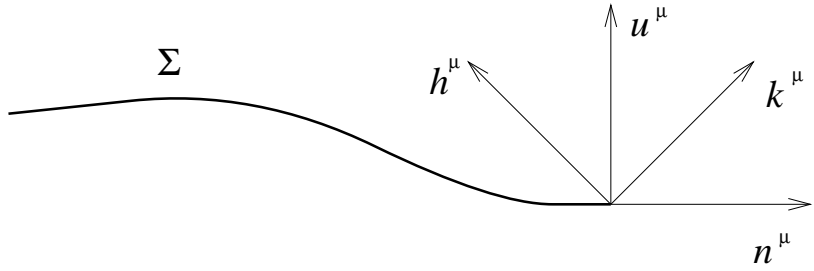,height=3.5cm}
\caption{Construction of the proper slicing}
\label{bondi2}
\end{figure}
Notice that the choice of $k^{\mu}$ will ensure that the energy that we compute
is associated with a proper rest frame such that the timelike normal behaves 
like $\tilde{u}^{\mu} \partial /\partial x^{\mu}=l^{-\frac{1}{2}}\partial 
/\partial v=N^{-1}\partial /\partial t$, which is obtained by using (\ref{ef2gs}).
Therefore, with $N=\sqrt{l(u)}$ we get
\begin{equation}
\label{bondiN}
N\left(\tilde{n}[\phi]-\tilde{n}[\phi]|^{0}\right)=
\left( l(u)-\sqrt{l(u)l^0(u)} \right)
\frac{\partial \phi}{\partial u}
\end{equation}  
and
\begin{equation}
\label{modifiedN}
N\tilde{n}[\phi]-\left(N\tilde{n}[\phi]\right)|^{0}=\left( l(u)-l^0(u) \right)
\frac{\partial \phi}{\partial u}.
\end{equation} 
Using (\ref{bondiN}) in (\ref{energy}) and remembering that 
$g(\phi)d\phi=du$ and $l(u)=l(\phi)g(\phi)$
we finally obtain
\begin{equation}
E_{BS}=4e^{-2\phi} \left( l(\phi)-\sqrt{l(\phi)l^{0}(\phi)} \right)
\end{equation} 
which is exactly the same as expression (\ref{fenergy}) and therefore the ADM 
mass and the Bondi-Sachs mass turn out to have the same value.
This can be readily understood by recalling that the difference between
the BS and the ADM mass is the integral of a stress energy flux which vanishes
at this purely classical level. 
Similarly (\ref{modifiedN}) exactly reproduces the result of (\ref{newE})
thereby showing that also the modified definition of mass on ${\cal{I^+}}$
agrees with the value at spatial infinity.

\section{Hawking radiation of generalized Schwarzschild
black holes}
There are a number of ways of calculating the Hawking radiation
\cite{rev}. One of them consists in comparing  vacua before
and after the formation of a black hole. In the case of generalized
dilaton gravity this way is technically rather involved. We
prefer a simpler approach based on an analysis of static black
hole solutions.

Consider a generalized Schwarzschild black hole given by
\begin{equation}
ds^2=-L(U)d\tau^2+L(U)^{-1}dU^2 , \label{gSch}
\end{equation}
where $L(U)$ has a fixed behavior at the asymptotic region ${\cal I}^+$:
\begin{equation}
L(U) \to L_0(U) \label{Jas}
\end{equation}
with $L_0(U)$ corresponding to the ground state solution.
At the horizon we have $L(U_h)=0$. We can calculate the geometric Hawking
temperature as the normal derivative of the norm of the Killing vector
$\partial /\partial \tau$ at the (nondegenerate) horizon \cite{wald2}
\begin{equation}
T_H=|\frac 12 L'(U_h) |. \label{THgen}
\end{equation}

Let us introduce the coordinate $z$ which is an analog of the
Regge-Wheeler {\it tortoise coordinate} \cite{wheeler}
$r^*$ for the ordinary Schwarzschild black hole
\begin{equation}
dU =dz L(U). \label{Jdefz}
\end{equation}
Then the metric takes the conformaly trivial form with
\begin{equation}
ds^2 = e^{2\rho} (-d\tau^2+dz^2) \qquad \rho =\frac 12 \ln L. \label{Jdefrho}
\end{equation}
In conformal coordinates the stress energy tensor looks
like \cite{rev}
\begin{equation}
T_{--}=-\frac 1{12} ((\partial_-\rho )^2 -
\partial_-\partial_-\rho )+t_- =T_{--}[\rho (L)]+t_-.
\label{JTmm}
\end{equation}
One can choose coordinates such that in the asymptotic
region
\begin{equation}
T_{--}[\rho (L_0)]=0 . \label{TJ0}
\end{equation}
This choice ensures that
there is no radiation in the ground state. It means
that we measure Hawking radiation of a black hole without any contribution
from background Unruh radiation.

The constant $t_-$ is defined by the condition at the horizon
\begin{equation}
T_{--}|_{\rm hor}=0  \label{Thor}
\end{equation}
in the spirit of  . \cite{CrFu}. 
The corresponding vacuum state is called the Unruh vacuum.
In this state there is no energy flux at the black hole horizon.
Of course, there cannot be such a thing as an observer at
the horizon. However, as we shall demonstrate bellow, predictions of the 
theory with regard to measurements made at infinity are independent 
of the choice of coordinates at the horizon. In the case of four
dimensional black holes this was shown in \cite{WaldCMP}.
Taking into account equations
(\ref{JTmm}), (\ref{TJ0}) and (\ref{Thor}) one obtains a relation
between $T_{--}[\rho ]$ at the horizon and the asymptotic value
of $T_{--}$:
\begin{equation}
T_{--}|_{\rm asymp}=-T_{--}[\rho ]|_{\rm hor} \label{ashor}
\end{equation}
The following simple identities are useful:
\begin{equation}
\partial_-=\frac 12 \partial_z ,\quad
\partial_z \rho = \frac 12 L' , \quad
\partial_z^2 \rho = \frac 12 L'' L ,
\label{useful}
\end{equation}
where prime denotes differentiation of $L$ with respect to
$U$. By substituting (\ref{useful}) into (\ref{ashor}) we obtain
the Hawking flux
\begin{equation}
T_{--}|_{\rm asymp}=\frac 1{48} (\frac 12 L'(U))^2|_{hor}.
\label{Jflux}
\end{equation}

It is easy to demonstrate that our result is independent of a particular
choice of conformal coordinates provided the behavior at the asymptotic
region is fixed. Without destroying the conformal gauge one can
change coordinates so that
\begin{equation}
\rho \to \rho +h_+(x^+)+h_-(x^-)
\end{equation}
with two arbitrary functions $h_\pm$. Since the behavior of $\rho$
at ${\cal I}^+$ is fixed, $h_-=0$. The transformation with arbitrary
$h_+$ does not change $T_{--}$.

Before we start to apply (\ref{Jflux}) to some specific space-times
we remark that as in the case of the ADM and Bondi mass the value will once
again depend on our choice of  the asymptotic behavior of the metric. 
We will use the same form of the metric as for the calculation of the ADM mass,
equations (\ref{pres}) and   (\ref{L}), which give the metric (\ref{gSch})
after a transformation like (\ref{ef2gs}).
For the most interesting space-times of 
our action (\ref{ldil}) we will explicitly present the solutions. \newline
\labelitemi $\quad$ {\bf Asymptotic Minkowski spacetimes:} \newline
As mentioned above  this corresponds to $b=a-1$. From (\ref{L}) we get for 
$a \neq 1$ 
\begin{equation}
L(U)=B_{1}B_{2}^{\frac{2-a}{2(a-1)}}U^{\frac{a}{a-1}}-1.
\end{equation}
The horizon determined by $L(U_h)=0$ is located at
\begin{equation}
U_h=B_{1}^{\frac{1-a}{a}}B_{2}^{\frac{a-2}{2a}}
\end{equation}
and therefore we get
\begin{equation}
T_{--}|_{\rm asymp}=\frac{a^2}{384} 
C^{\frac{2(a-1)}{a}}\left( \frac{2B}{a} \right)^{\frac{2-a}{a}}.
\end{equation}
For the special case $a=1$, the CGHS model, we have
\begin{equation}
L(U)=\frac{e^{\sqrt{B}U}C}{2B}-1
\end{equation}
which results in
\begin{equation}
T_{--}|_{\rm asymp}=\frac{\lambda^2}{48},
\end{equation}
thereby confirming the well known result for the CGHS model, where we used the
identification $B=4\lambda^2$ such that (\ref{ldil}) reproduces the
CGHS action.
For the case of \newline
\labelitemi $\quad$ {\bf Asymptotic de-Sitter spacetimes} \newline
we have from (\ref{desit}) 
\begin{equation}
L(U)=B_1B_2^{\frac{2-a}{2(a-1)}}U^{\frac{a}{a-1}}-B_2U^2,
\end{equation}
and get for the value of Hawking radiation
\begin{equation}
T_{--}|_{\rm asymp}=\frac{2-a}{192(a-1)} 
\left(B_1^{1-a}B_2^{\frac a2} \right)^{\frac{1}{2-a}}.
\end{equation}
Setting $b=0$ we have \newline
\labelitemi $\quad$ {\bf Asymptotic Rindler spacetime.}
According to (\ref{L}) we have
\begin{equation}
L(U)=B_1B_2^{\frac{2-a}{2(a-1)}}U^{\frac{a}{a-1}}-B_2^{\frac 12}U
\end{equation}
and the Hawking radiation reads
\begin{equation}
T_{--}|_{\rm asymp}=\frac{B}{384}\frac{1}{(a-1)}.
\end{equation}
Notice, that if we had not used the rescaling coefficient as in (\ref{L}), but
if instead we had set
\begin{equation}
K=\frac{B}{4},
\end{equation}
we would have obtained
\begin{equation} 
T_{--}|_{\rm asymp}=\frac{\lambda^2}{48},
\end{equation} 
which is just the result given in \cite{fabbri}.

\section{Path integral measure, asymptotic conditions
and the problem of interpretation}

It is well known (see e.g. \cite{Fuj}) that there is a unique
ultralocal path integral measure for a scalar field $f$ yielding a
covariantly conserved stress energy tensor. This measure is
defined by the equation
\begin{equation}
\int {\cal D} f \exp \left ( i\int \sqrt{-g} d^n x f^2
\right ) =1 .
\label{measure}
\end{equation}
It is also known \cite{LDF} that in four dimensional Einstein
gravity one can define energy in a unique and self-consistent way
if and only if in the asymptotic region the metric is approaching fast enough
that of flat Minkowski space. As a consequence, Hawking
radiation and ADM mass depend on a choice of asymptotic
conditions and path integral measure.

In dilaton gravity we have a new entity
which is absent in Einstein gravity, the dilaton
field. One is tempted to use a rescaled metric at some steps
of the calculations. In the present paper such a rescaling is trivial
everywhere. This means that we are using just the metric $g_{\mu\nu}$
which appears in the action (\ref{ldil})
to define the path integral measure, stress energy tensor etc.
Of course, this is just a matter of interpretation. One can claim
that the rescaled metric $\Phi (\phi )g_{\mu \nu}$ describes the geometry of
space-time and thus should be used in the definitions of the above mentioned
quantities.  However, the rescaled metric should be used everywhere.
Otherwise, the quantum theory becomes inconsistent. For example, a part
of diffeomorphism invariance can be lost. We believe that if one wishes
to keep contact to four dimensional Einstein gravity, which is
diffeomorphism invariant, one should retain general coordinate
invariance\footnote{Some new possibilities appear if one trades
diffeomorphism invariance for Weyl invariance \cite{Weyl}.
We shall not consider such theories here.}.
In the context of dilaton gravity models a transition from $g_{\mu\nu}$
to $g^\Phi_{\mu\nu}=\Phi (\phi )g_{\mu \nu}$ means replacement of one
particular dilaton interaction by another. Such a replacement should in
general change the Hawking radiation because the definitions of the stress
energy tensor and path integral measure and the asymptotic behavior of
metric are not conformally invariant. At least, conformal invariance
of the Hawking radiation was never proved. Such conformal equivalence
was conjectured in a recent paper by Cadoni \cite{Cadoni} who used
conformal transformation in general dilaton theory to remove kinetic
term $(\nabla\phi)^2$ of the dilaton field. He also used a
$\phi$-dependent path integral measure and $\phi$-dependent stress
energy tensor to define the Hawking radiation. 
A comparison shows that his results for the Hawking temperature differ by an
$a$ and $b$ dependent factor from ours, which were obtained
directly without use of conformal transformation. We conclude that
conformally equivalent theories do really give different results
for the Hawking radiation. This statement can be illustrated by an
example from .\cite{CaMi}, where the CGHS black hole was
transformed to the flat Rindler space-time. Due to the absence of
black hole curvature singularity any radiation in the latter space
should be considered as the Unruh radiation rather than the Hawking one.

We have seen that both ADM mass and Hawking radiation depend
on the asymptotic behavior of the metric and on the subtraction procedure
of reference space-time contribution. This dependence is quite natural
from a physical point of view. Since energy and energy flux are not
coordinate invariant, in addition to a zero energy state one should also
define two observers. One of them corresponds to a particular
solution, and the other to the reference zero point configuration.
For asymptotically Minkowski, Rindler or de Sitter solution there
is a choice which is clearly preferable. With this choice our results
for ADM mass and Hawking radiation are independent of coordinate
transformations which vanish rapidly enough at asymptotic region.
In the case of four dimensional Einstein gravity appropriate
asymptotic conditions were formulated by Faddeev \cite{LDF}.
For the CGHS model with Polyakov--Liouville term the suitable asymptotic
conditions were found recently \cite{kim}.

\section{Dynamical formation of black holes}
\subsection{General solution}

Consider a scalar matter field $f$ minimally coupled to gravity.
We add the following term to the action (\ref{ldil})
\begin{equation}
L_m= -\frac 12 \int d^2x \sqrt{-g} g^{\mu\nu} \nabla_\mu f
\nabla_\nu f . \label{lmat}
\end{equation}
In the presence of this matter field the equations of motion
take the form
\begin{eqnarray}
g_{\mu \nu}((4-2a)(\nabla \phi )^2-2\nabla^2 \phi -\frac B2
e^{2\phi (1-a-b)})+4(a-1)\partial_\mu \phi \partial_\nu \phi
& &\nonumber \\
+2\nabla_\mu \nabla_\nu \phi+\frac 12 e^{2\phi}T_{\mu\nu}
&=0& \nonumber\\
R-4a(\nabla \phi )^2 +4a\nabla^2\phi +B(a+b)
e^{2(1-a-b)\phi}&=0& \nonumber \\
\nabla^2 f&=0& \label{eqmo}
\end{eqnarray}
and in conformal gauge these equations are
\begin{eqnarray}
2\partial_+ \partial_- \rho +4a \partial_+ \phi \partial_- \phi -4a \partial_+ 
\partial_- \phi +
\frac B4 (a+b) e^{2(\rho +(1-a-b)\phi )} =0
\label{eqmo1} \\
-4\partial_+ \phi \partial_- \phi +2\partial_+ \partial_- \phi -
\frac B4 e^{2(\rho +(1-a-b)\phi )} =0
\label{eqmo2} \\
\partial_+ \partial_- f=0 \label{eqmo3} \\
4(a-1)\partial_\pm \phi \partial_\pm \phi +
2\partial_\pm \partial_\pm \phi -4\partial_\pm \rho \partial_\pm \phi
=\frac 12 e^{2\phi} \partial_\pm f \partial_\pm f. \label{eqmo4}
\end{eqnarray}

The matter equation of motion (\ref{eqmo3}) can be
solved by setting $f=f(x^+)$. Then $\rho$ can be
determined from the $(-,-)$ component of 
(\ref{eqmo4}):
\begin{equation}
\partial_- \rho =(a-1) \partial_- \phi +\frac 12 \partial_- \ln |\partial_- \phi |
\end{equation}
which gives
\begin{equation}
\rho =(a-1) \phi +\frac 12 \ln |\partial_- \phi | +\xi (x^+ )
\label{rhop}
\end{equation}
The arbitrary function $\xi (x^+ )$ can be removed by using
residual gauge freedom of the conformal gauge. In what
follows we set $\xi =0$.
$\rho$ can be removed from the equations of motion which
are reduced to the following three independent equations:
\begin{eqnarray}
\frac 12 (\partial_+ \phi ) \partial_+ [\ln |\partial_+ 
\phi |-\ln |\partial_- \phi | ]
=\frac {e^{2\phi}}8 (f')^2 \label{eq1} \\
\partial_+ \partial_- (\phi -\frac 12 \ln |\partial_- \phi |)=
\frac {Bb}8 |\partial_- \phi |e^{-2b\phi} \label{eq2} \\
-2\partial_+ \partial_- \phi +4\partial_+
 \phi \partial_- \phi +\frac B4 |\partial_- \phi |
e^{-2b\phi} =0 \label{eq3}
\end{eqnarray}
After some algebra the eq. (\ref{eq2}) gives
\begin{equation}
\partial_+ e^{-2\phi} \mp \frac B{8(b+1)} e^{-2\phi (b+1)}=
-2\eta (x^+)e^{-2\phi} +\lambda (x^+)
\label{eq2.}
\end{equation}
with two arbitrary functions $\eta$ and $\lambda$. From eq.
(\ref{eq3}) one obtains
\begin{equation}
\partial_+ e^{-2\phi} \mp \frac B{8(b+1)} e^{-2\phi (b+1)}=
\kappa (x^+) \label{eq3.}
\end{equation}
This means that $\eta =0$, $\kappa =\lambda$ and
\begin{equation}
\partial_+ \phi =\mp \frac B{16(b+1)} e^{-2b\phi} -\frac 12 \lambda
(x^+) e^{2\phi}
\label{eq23}
\end{equation}
Here $\pm$ is the sign of $\partial_- \phi$. From (\ref{eq1}) we can
express the arbitrary function $\lambda$:
\begin{equation}
\partial_+ \lambda (x^+)=-\frac 12 (f')^2
\label{eq1.}
\end{equation}
The two equations (\ref{eq23}) and (\ref{eq1.})
completely define $\phi$ for a given matter distribution.
Let us take the matter field in the form of shock wave
\begin{equation}
\frac 12 f'(x^+)^2=D\delta (x^+-x^+_0)
\label{shock}
\end{equation}
Before the shock wave starts we have the empty space solution:
\begin{eqnarray}
\phi =\frac 1{2b} \ln \vert \frac {Bbt}{4(b+1)} \vert ,
& {\rm for}& \frac B{b+1} >0 \nonumber \\
\phi =\frac 1{2b} \ln \vert \frac {Bbx}{4(b+1)} \vert ,
& {\rm for}& \frac B{b+1} <0 \label{flatsol}
\end{eqnarray}
Here we consider only the case $b\ne 0$, since the dynamical
formation of black holes for $b=0$ was already considered in \cite{fabbri}.
After the shock wave the black hole solution is formally given
by the integral
\begin{equation}
\int^{\phi (x,t)} {{de^{-2\phi}}\over{\pm \frac B{8(b+1)}e^{-2(b+1)\phi}
-D}} =x^+ +h(x^-) \label{soll}
\end{equation}
Now we should glue together the
solutions (\ref{flatsol}) and (\ref{soll}) by using the arbitrary
function $h(x^-)$. For the sake of simplicity we put $x^+_0=0$. Note,
that we can restrict ourselves to positive or negative values of
coordinates $x$ and $t$ in (\ref{flatsol}) since in these regions the
dilaton field changes from $-\infty$ to $+\infty$. On the line $x^+=0$
we obtain from (\ref{flatsol})
\begin{equation}
\phi_{bou}(x^-)= \frac 1{2b} \ln \vert \frac {Bbx^-}{8(b+1)} \vert .
\end{equation}
Next one can find the function $h(x^-)$ which ensures continuity
of $\phi$:
\begin{equation}
h(x^-)=\int^{\phi_{bou}(x^-)}
{{de^{-2\phi}}\over{\pm \frac B{8(b+1)}e^{-2(b+1)\phi} -D}}
\label{hofx}
\end{equation}
The lower limit in the integrals in (\ref{soll}) and (\ref{hofx})
plays no role, however it should be the same and constant in both
equations. The $\pm$ sign in (\ref{soll}) can be fixed by calculating
$\partial_-\phi$.

\subsection{Asymptotically Minkowski solutions}

Consider now the case $a=b+1$ corresponding to asymptotically
Minkowski solutions in more detail. Let $0<a<2$. One can easily
demonstrate that in the asymptotic regions before and after the
formation of a black hole the metric of the solutions (\ref{soll})
and (\ref{flatsol}) behaves as
\begin{equation}
e^{2\rho} \to |B/16a| .\label{asrho}
\end{equation}
Thus to obtain a black hole surrounded by
Minkowski space with unit metric we need a proper rescaling of
coordinates.

Let us modify the procedure of the previous subsection in order to
obtain the rescaled solution. First of all, one should take the
function $\xi$ in (\ref{rhop}) in the form
\begin{equation}
\xi =\frac 14 \ln \left| \frac {16a}{B} \right| .
\label{newxi}
\end{equation}
With this choice equation (\ref{rhop}) now reads
\begin{equation}
\rho = (a-1)\phi +\frac 12 
\ln \vert 4 \sqrt{\left| \frac aB \right|} \partial_- \phi \vert . \label{newrhop}
\end{equation}
One can repeat all the steps of the previous subsection with
this form of $\rho$. As a result, one obtains a unit metric
for $x^+<0$ and an asymptotically unit metric for $x^+>0$.
Such a configuration is again produced by the shock wave
(\ref{shock}), where the constant $C$ is now related to the
parameter $D$ in a different manner,
\begin{equation} 
D=\frac C4 \left( \frac{a}{B} \right)^{\frac 12}. \label{DC}
\end{equation} 
This constant is proportional to the energy of the shock wave
measured by a Minkowski space observer. This confirms our previous
result (\ref{admmink}) for the ADM mass of black hole in asymptotically
minkowskian space-time.

It is tempting to calculate the Hawking radiation by comparing
normal ordering prescriptions and Green functions in two
asymptotic regions, before and after formation of a black hole
\cite{reviews}. If such a method could be applied in the coordinate
systems $(x^+,x^-)$ and $(x^+,h(x^-))$, the stress energy tensor
of the Hawking radiation would be calculated by differentiating
the function $h(x^-)$. This appeared, however, not to be the case.
The full ground state solution is covered by positive (or negative)
values of $x$ or $t$. Hence the Green function differs from the standard
expression valid for infinite range of all coordinates.

\section{Conclusions}
In the present paper we calculated the ADM and BS mass and Hawking radiation
of the black hole solutions in generalized dilaton gravities
described by the action (\ref{ldil}). Special attention was paid to
asymptotic conditions for the metric field. It turned out that both
mass and stress energy tensor depend on these conditions. From
a physical point of view this means that they are influenced by the choice
of an asymptotic observer. For generic $a$ and $b$ this choice is not
unique. It rather depends on the interpretation of particular two
dimensional model and its relation to four dimensional gravity.
For the selected values of $a$ and $b$ corresponding to asymptotically
Minkowski, Rindler or de Sitter solutions there exists a preferred
coordinate frame at infinity. This fact allows some
physical conclusions. We can see that, depending on the parameters
of the dilatonic action, the Hawking temperature can be related to the black
hole mass with positive or negative power. In a fully quantized
theory the parameters $a$ and $b$ should become running couplings
being functions of a scale parameter which
could be the black hole mass. Thus it could happen that a black
hole which starts its evolution near the point $a=-b=\frac 12$,
corresponding to spherically symmetric gravity, migrates in 
the course of evaporation to another region of the $(a,b)$ plane
where the heat capacity becomes positive. With our present level of 
understanding the last statement is nothing  more than an unsupported
speculation, which should, however, be able to describe black hole evolution in
some realistic quantum theory.

\section*{Acknowledgement}

We are grateful to W. Kummer for important suggestions.
Especially we thank S.R. Lau for advice on the subtleties of quasilocal energy.
We also thank L.D. Faddeev for enlightening remarks 
on the definition of energy in general relativity.
This work has been supported by Fonds zur F\"orderung der
wissenschaftlichen For\-schung (FWF) Project No.\ P 10221--PHY.  One
of the authors (D.V.) thanks GRACENAS for financial support.

\end{document}